\documentclass[3p,times,procedia]{elsarticle}
\usepackage{nupha_ecrc}
\usepackage{lineno}

\volume{00}
\firstpage{1}
\journalname{Nuclear Physics A}
\runauth{V. Klochkov and I. Selyuzhenkov for the NA61/SHINE Collaboration}
\jid{nupha}
\jnltitlelogo{Nuclear Physics A}
\usepackage{amssymb}
\usepackage[figuresright]{rotating}
\begin{document}
\begin{frontmatter}
\dochead{XXVIIth International Conference on Ultrarelativistic Nucleus-Nucleus Collisions\\ (Quark Matter 2018)}
\title{NA61/SHINE​ ​measurements​ ​of​​ anisotropic​​ flow​ relative​​ to​​ the​ ​spectator plane​ ​in​ ​Pb+Pb​ ​collisions​ at 30$A$ GeV/$c$}
\author{V. Klochkov$^{\mathrm{a,b}}$ and I. Selyuzhenkov$^{\mathrm{a,c}}$ for the NA61/SHINE Collaboration}
\address[1]{GSI Helmholtzzentrum f\"ur Schwerionenforschung, Darmstadt, Germany}
\address[2]{Goethe-University Frankfurt, Frankfurt, Germany}
\address[3]{National Research Nuclear University (Moscow Engineering Physics Institute), Moscow, Russia}

\begin{abstract}
We present an analysis of the anisotropic flow harmonics in Pb+Pb collisions at beam momenta of 30$A$ GeV/$c$ collected by the NA61/SHINE experiment in the year 2016. Directed and elliptic flow coefficients are measured relative to the spectator plane estimated with the Projectile Spectators Detector (PSD). The flow coefficients are reported as a function of transverse momentum in different classes of collision centrality. The results are compared with a new analysis of the NA49 data for Pb+Pb collisions at 40$A$ GeV using forward calorimeters (VCal and RCal) for event plane estimation.
\end{abstract}

\end{frontmatter}

\section{Introduction}
\label{}

The NA61/SHINE experiment at the CERN SPS recently extended its program for the energy scan with Pb ions in the energy range of 13-150$A$ GeV/$c$. Compared to the existing data from the NA49 experiment~\cite{Alt:2003ab} at the CERN SPS, the new data allows for more precise measurements of anisotropic flow harmonics. The fixed target setup of NA61/SHINE also allows one to extend the flow measurements carried out by the STAR Collaboration during the RHIC beam energy scan (BES) program~\cite{PhysRevLett.112.162301} to a wide rapidity range, extending up to the forward region where projectile nucleon spectators appear. The NA61/SHINE measurements with Pb ions and the experimental techniques using spectators at the lowest energy available at the SPS are also relevant for studies of collisions at energies of few GeV per nucleon~\cite{KARDAN2017812}, in particular for the preparation of the Compressed Baryonic Matter (CBM) heavy-ion​ experiment​ at​ the​ future​ FAIR​ facility​ in​ Darmstadt.

\section{Data sample and analysis setup}
\label{}
A sample of Pb+Pb collisions at 30$A$ GeV/$c$ was used for the analysis. Events with fitted vertex position close to the target region were selected. Events overlapping in time (pileup) were rejected. After event selection, the available statistics is 1.1~M events for the minimum bias trigger (T4) and 0.6~M events for the central trigger (T2) which was fully efficient in the 0-15\% centrality class.

In the NA61/SHINE experiment, the momentum measurement and the identification of charged pions and protons are done with the Time Projection Chambers (TPC). Charged particle tracks with total number of clusters in the TPCs larger than 30 and number of clusters in the vertex TPCs larger than 15 were accepted for flow analysis. To avoid track splitting, the number of hits associated to the track was required to be more than 55\% of the maximum number of points along the particle trajectory. Primary tracks were selected based on the distance of closest approach to the primary vertex in the plane transverse to the beam direction, which was required to be less than 2 cm in $x$ direction and less than 1 cm in $y$ direction. Charged pion and proton identification was based on specific energy loss dE/dx in the TPCs.

\begin{figure}[ht]
\centering
\includegraphics[width=0.44\textwidth] {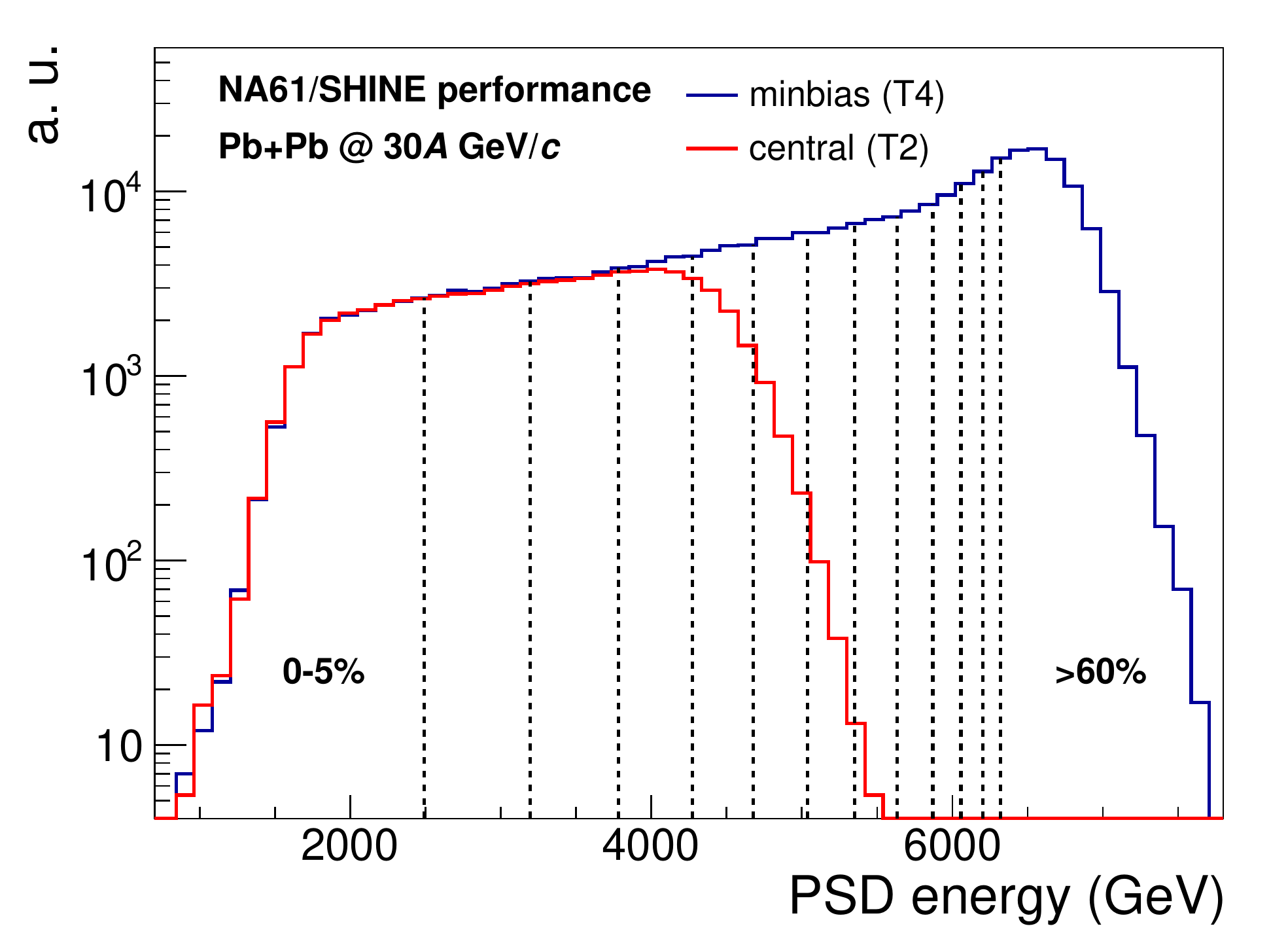}
\includegraphics[width=0.44\textwidth] {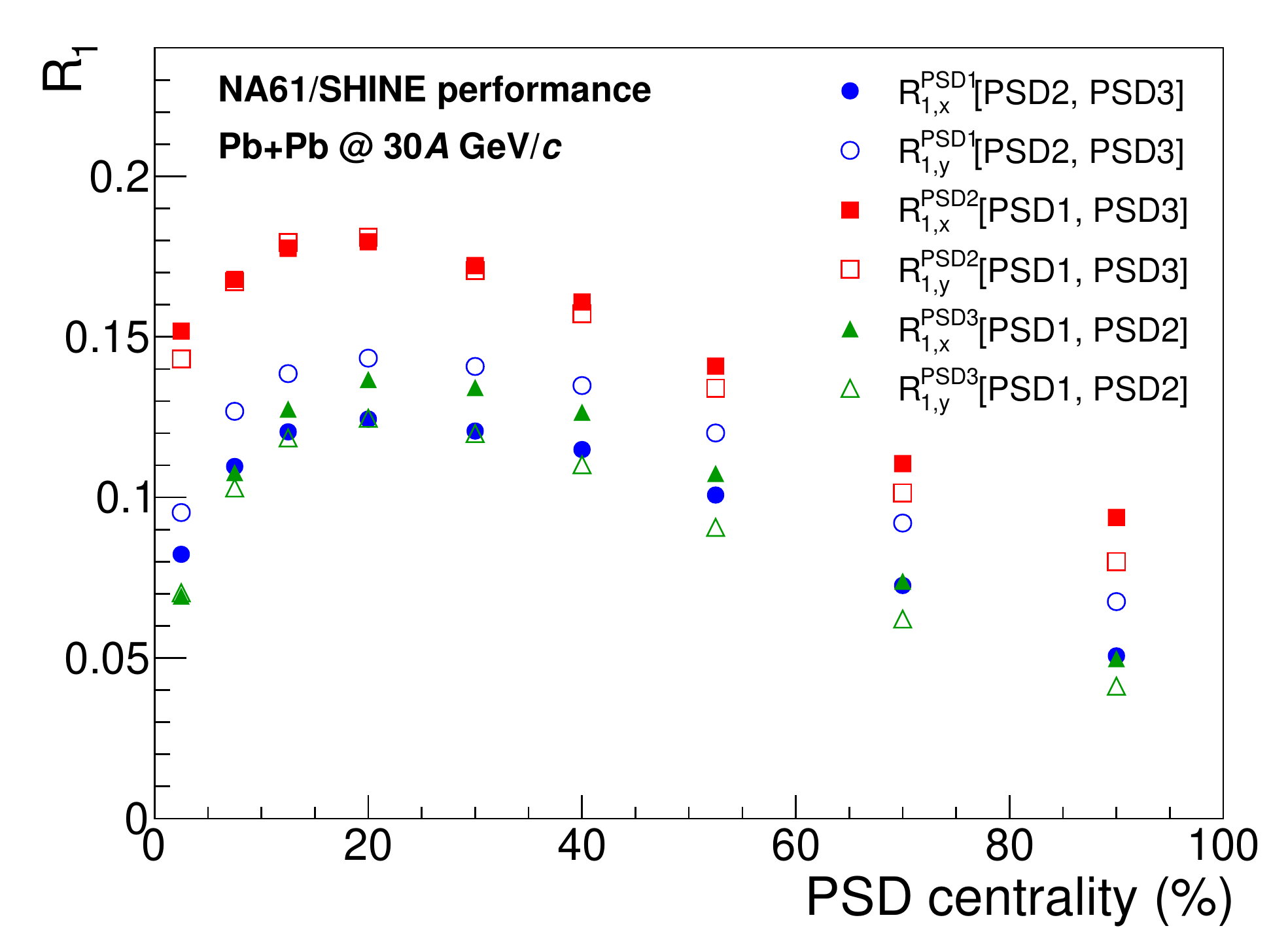}
\caption{
Left: PSD energy distribution for central (T2) and minimum bias (T4) triggers. Dashed vertical lines mark the borders of event (centrality) classes.
 Right: Resolution correction factors $R_{1,\alpha}^A\{B,C\}$ obtained for PSD subevents using the 3-subevent method. 
}
\label{fig:centr_res}
\end{figure}

The Projectile Spectator Detector (PSD) is sensitive mostly to spectator fragments (outer modules are also sensitive to produced particles) and is used for event (centrality) classification and reaction plane determination.
Event classification is performed following the procedure described in Ref.~\cite{Klochkov:2016GSI}. 
Figure~\ref{fig:centr_res}~\textit(left) shows the result of the event classification procedure using the energy measured at forward rapidity with the PSD.

To estimate the reaction plane orientation it is common to use the azimuthal asymmetry of particle production in the plane transverse to the beam direction. Due to the momentum transfer between participants and spectators, the spectators  (fragments of projectile and target nuclei) are deflected in the course of the collision.
For non-central collisions, the asymmetry of the initial energy density  in the transverse plane is expected to be aligned with the direction of the reaction plane, and thus the spectator deflection direction is likely to be correlated with the impact parameter (or reaction plane) direction.
One can estimate the reaction plane angle with spectators detected in the PSD and determine the flow of produced particles detected in the TPCs with respect to this plane. For event plane determination PSD modules were subdivided into 3 groups (PSD1, PSD2, PSD3) with approximate coverage in pseudorapidity $\eta \in \{ (5.1, \infty), (4.4, 5.1), (4.0, 4.4) \}$.
The azimuthal asymmetry of the measured distributions is described in terms of two dimensional flow vectors ${\textbf q_n}$ and ${\textbf Q_1}$ determined event-by-event from the TPC tracks and PSD subevents:
\begin{equation}
{\textbf Q_1^{s}} = \frac{1}{E_{s}}\sum_{j}^{N_s}~E_{j}~{\textbf n_{j}}~;~~
{\textbf q_n} = \frac{1}{M}\sum_i^{M}{\textbf u_{n,i}}~;~~
{\textbf u_{n,i}} = \{\cos n \varphi_i,\sin n \varphi_i\}~,
\label{Eq:Q_vectors}
\end{equation}
where the unit vector ${\textbf n_{j}}$ points to the center of the $j$-th PSD module, $E_{j}$ is the energy deposition in the $j$-th module, $N_s$ is the number of modules and $E_{s} = \sum_{j}^{N_s}~E_{j}$ is the total energy of the PSD subevent $s$=\{PSD1,~PSD2,~PSD3\}.
For each particle track $i$ reconstructed with the TPC a $n$-th harmonic unit vector ${\textbf u_{n,i}}$ is defined.
The ${\textbf q_n}$-vectors were calculated for charged pions and protons in transverse momentum ($p_{T}$) intervals using Eq.~(\ref{Eq:Q_vectors})
where $M$ is the number of particles in a given $p_{T}$ interval.

Independent estimates of the flow harmonics $v_{n}$ can be obtained using the scalar product method:
\begin{equation}
v_{1}^{\alpha}\{A;B,C\}=\frac{2\langle q_{1,\alpha}Q_{1,\alpha}^{A}\rangle}{R_{1,\alpha}^A\{B,C\}}~;~~
v_{2}^{\alpha \beta \gamma}\{A,B;C\}=\kappa_{\alpha \beta \gamma} \frac{4\langle q_{2,\alpha}Q_{1,\beta}^{A}Q_{1,\gamma}^{B}\rangle}{R_{1,\beta}^A\{B,C\} R_{1,\gamma}^B\{A,C\}},
\label{Eq:Flow_v12}
\end{equation}
where for $v_1$ $\alpha=x,y$, for $v_2$ $\kappa_{\alpha \beta \gamma}=1$ for $(\alpha, \beta, \gamma) = \{(x,x,x), (y,x,y), (y,y,x)\}$ and $\kappa_{\alpha \beta \gamma}=-1$ for $(\alpha, \beta, \gamma)=(x,y,y)$. ${\textbf Q}$-vector resolution correction factors $R_{1,\alpha}^A\{B,C\}$ are calculated using the 3-subevent method:
\begin{equation}
R_{1,\alpha}^A\{B,C\}=\sqrt{2\frac{\langle Q_{1,\alpha}^{A}Q_{1,\alpha}^{B}\rangle\langle Q_{1,\alpha}^{A}Q_{1,\alpha}^{C}\rangle}{\langle Q_{1,\alpha}^{B}Q_{1,\alpha}^{C}\rangle}}~.
\label{Eq:Flow_R}
\end{equation}

Imperfect acceptance and efficiency of the detector bias the azimuthal angle distribution of measured particles. Corrections for detector inefficiency as a function of transverse momentum and rapidity were determined using a Monte-Carlo GEANT4 simulation of the NA61/SHINE experiment with the DCM-QGSM heavy-ion event generator~\cite{dcmqgsm}. A flow vector correction procedure for detector anisotropy in azimuthal angle~\cite{Selyuzhenkov:2007zi} was implemented in a modular object-oriented C++ package QnCorrections framework~\cite{Gonzalez:2016GSI,QnGithub:2015}. Recentering, twist, and rescaling corrections were applied as a function of time and centrality. Figure~\ref{fig:centr_res}~\textit{(right)} shows the resolution correction factors $R_{1}$ for the $x$ and ${y}$ components of the three PSD subevents.
\section{Results}
\label{}
Results are presented for correlations between charged pions and protons produced by strong interaction processes and their weak and electromagnetic decays (in the TPC acceptance~\cite{tpc_acceptance}) and all hadrons at forward rapidity (in the PSD acceptance~\cite{psd_acceptance}). The results are corrected for detector non-uniformity. No corrections for secondary interactions were applied.
Results for $v_1$ and $v_2$ with different PSD subevents and different $x$ and $y$ components are found to be consistent within statistical uncertainties. The presented preliminary results are for all PSD subevents combined. For $v_1$ only the $x$-component correlation is used and for $v_2$ an average of the $(y,x,y)$ and $(y,y,x)$ correlations is used. These combinations have the smallest statistical uncertainties due to the geometry of the TPCs.  
\begin{figure}[ht]
\centering
\includegraphics[width=0.44\textwidth] {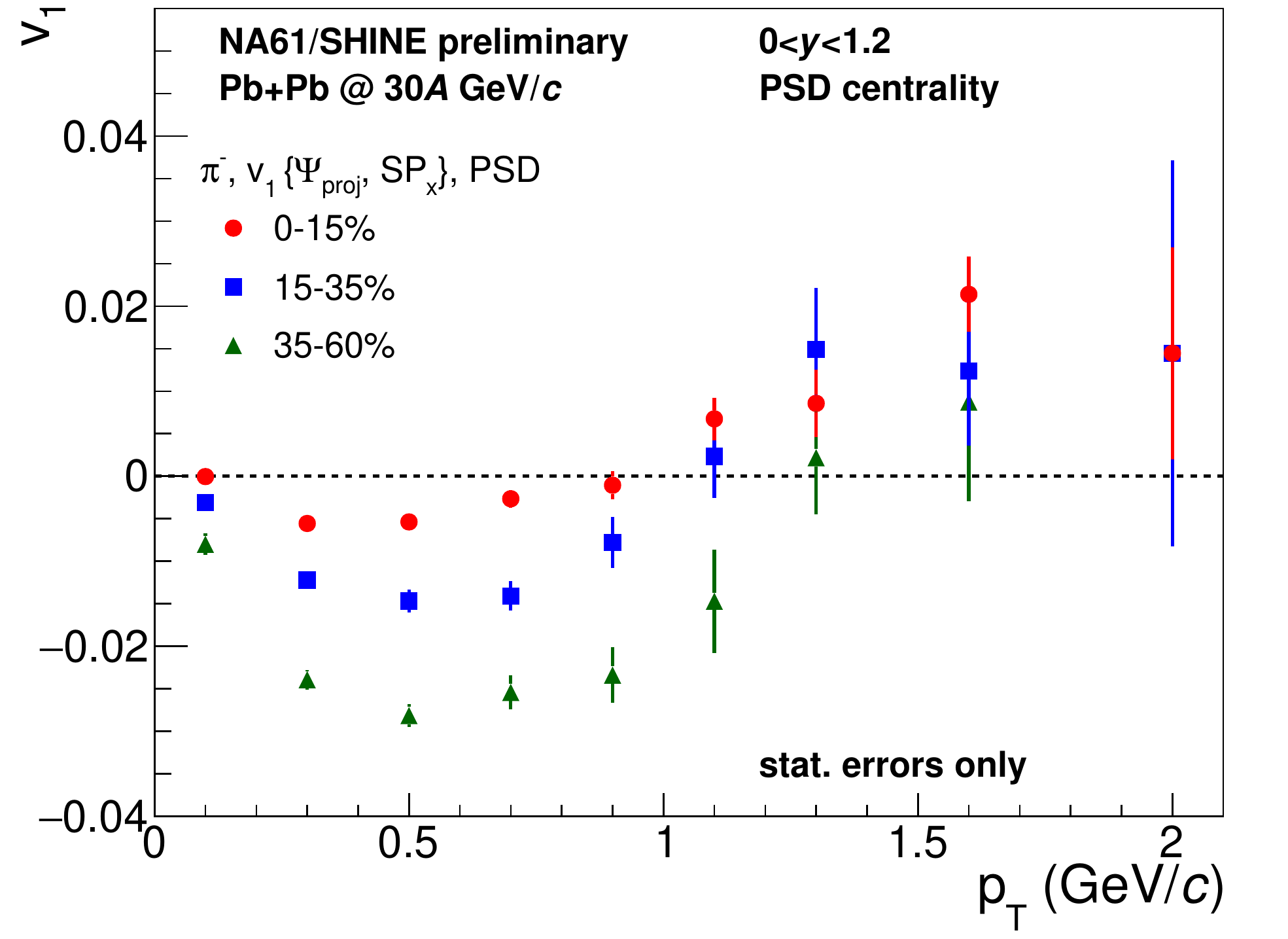}
\includegraphics[width=0.44\textwidth] {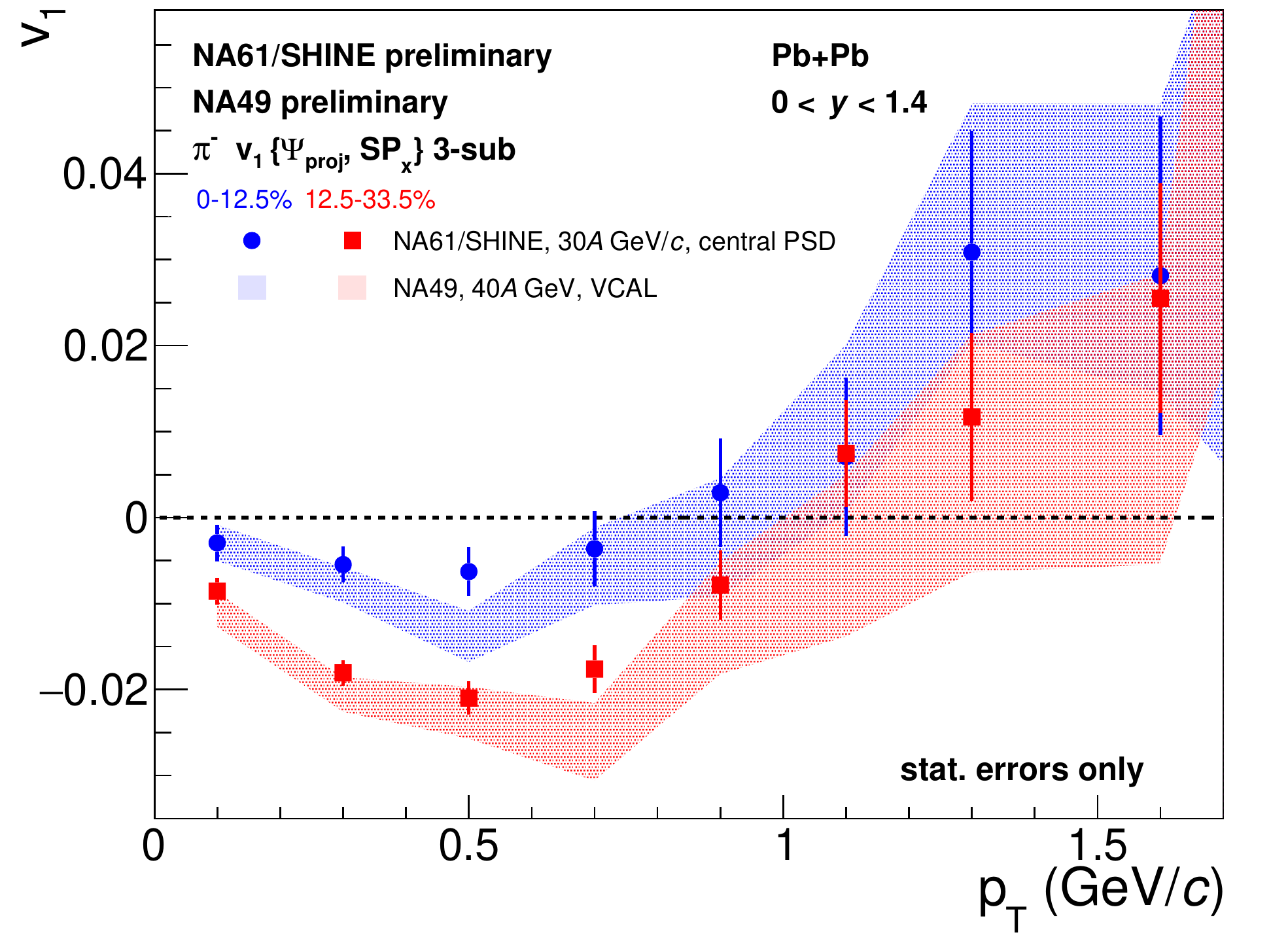}
\caption{
Left: negatively charged pion directed flow vs. transverse momentum ($p_T$) for different centrality classes.
Right: negatively charged pion directed flow compared to a new analysis~\cite{na49:poster} of the NA49 Pb+Pb data at 40$A$ GeV. The NA61/SHINE results are obtained using central PSD modules for centrality and projectile spectator plane ($\mathrm{\Psi_{proj}}$) estimation. Only statistical uncertainties are shown.
}
\label{fig:v1_all_centr}
\end{figure}
Figure~\ref{fig:v1_all_centr}~\textit{(left)} shows the directed flow of negatively charged pions as a function of transverse momentum in different centrality classes. 
A strong centrality dependence of $v_1$ is observed. At $p_T\approx$~0 directed flow approaches 0 and changes sign depending on centrality in the region $p_T\approx$ 0.9-1.3~GeV/$c$.
Figure~\ref{fig:v1_all_centr}~\textit{(right)} shows the $\pi^{-}$ directed flow as a function of $p_T$. It is compared to the new analysis~\cite{na49_poster} of NA49 data for Pb+Pb collisions at 40$A$ GeV using forward calorimeters (VCal and RCal) for projectile spectator plane estimation. Agreement between NA61/SHINE and NA49 measurements within statistical errors is observed.
\begin{figure}[ht]
\centering
\includegraphics[width=0.44\textwidth] {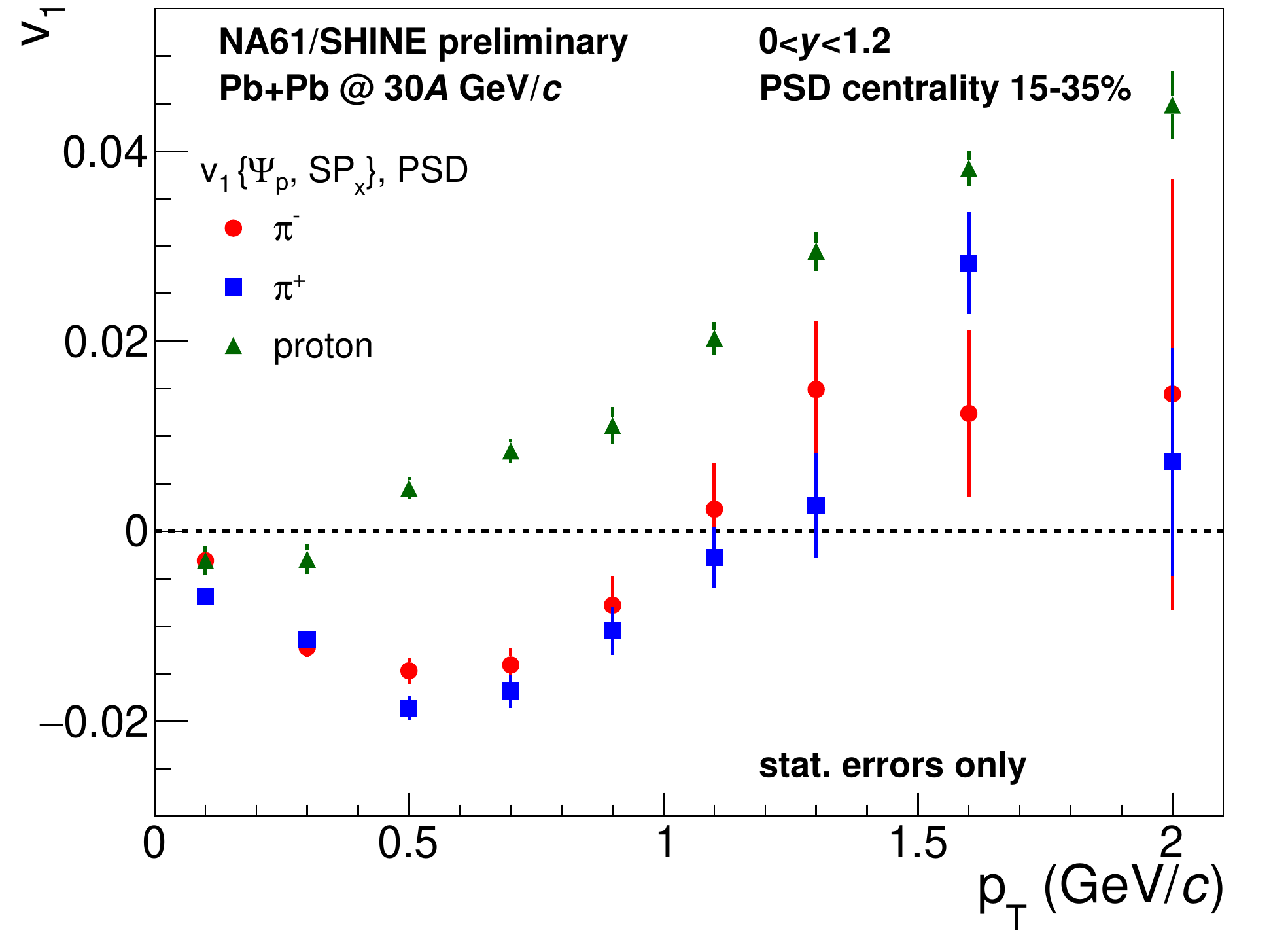}~~~%
\includegraphics[width=0.44\textwidth] {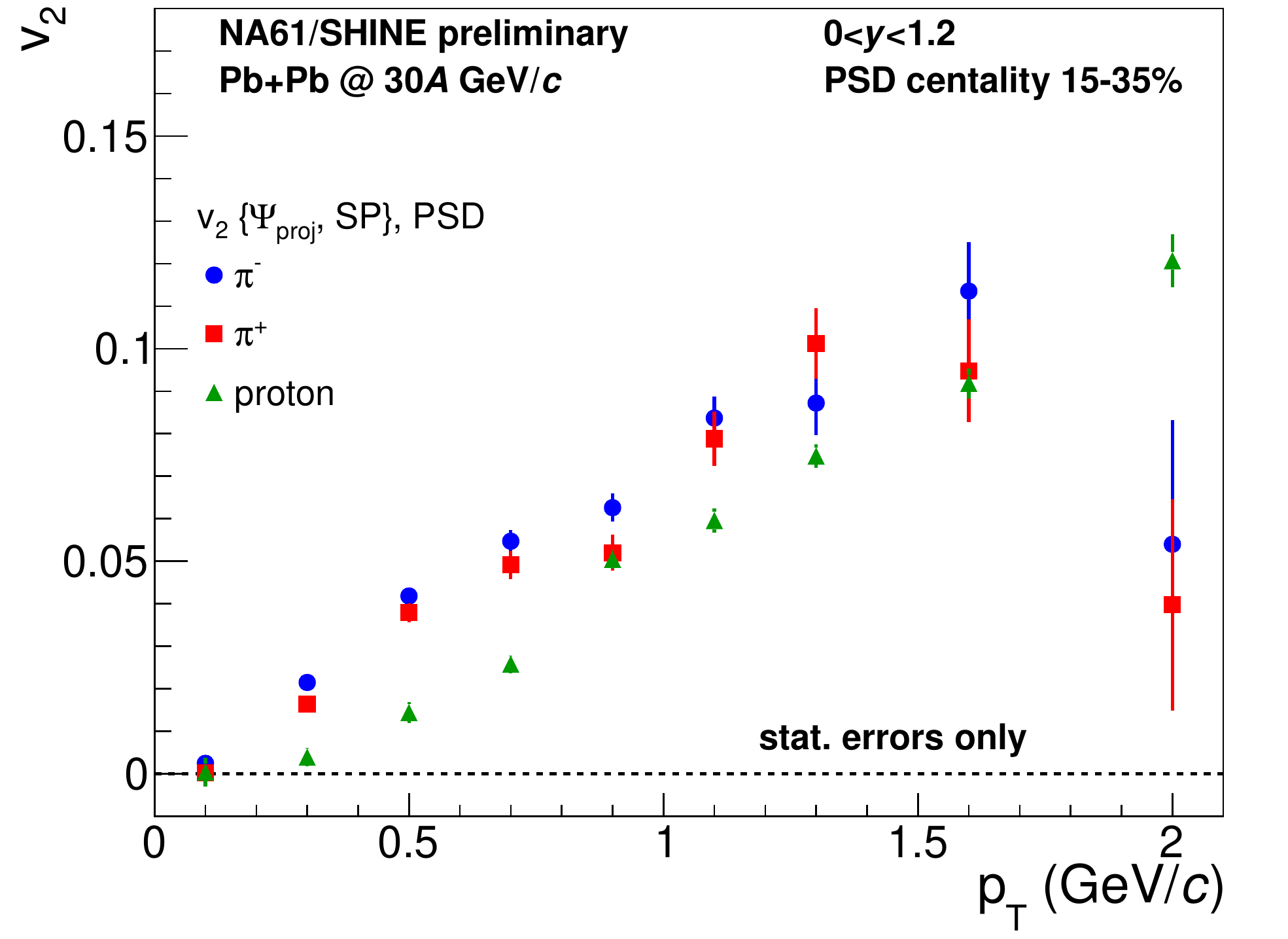}
\caption{
Charged pion and proton directed~\textit(left) and elliptic~\textit(right) flow as a function of transverse momentum for 15-35\% centrality class. Only statistical uncertainties are shown.
}
\label{fig:v1_v2_all_part}
\end{figure}
Results for $\pi^{-}$, $\pi^{+}$ and proton directed ($v_1$) and elliptic ($v_2$) flow for the 15-35\% centrality class are presented in Fig.~\ref{fig:v1_v2_all_part}. For both harmonics, a strong mass dependence is observed. The charge dependence of pion $v_1$ can be sensitive to effects of the magnetic field in heavy-ion collisions~\cite{Rybicki:2013qla}.
\section{Summary and Outlook}
\label{}
Preliminary results for charged pion and proton directed and elliptic flow in Pb+Pb collisions at 30$A$ GeV/$c$ recorded in 2016 by the NA61/SHINE experiment are presented for different centrality classes as a function of transverse momentum. Results are compared to a new and similar analysis~\cite{na49_poster} of the NA49 data using the forward calorimeters (VCal and RCal).
In future, measurements will be extended to other collision energies (13$A$ and 150$A$ GeV/$c$) and collision systems, such as Xe+La, Ar+Sc, and Be+Be collisions, which are available from the system size scan of the NA61/SHINE program. The developed measurement technique using spectators in a fixed target geometry is also relevant for physics performance studies of the future CBM experiment at FAIR at beam momentum of 12$A$ GeV/$c$ and below. CBM is constructing a projectile spectator detector with a design similar to that of the PSD of NA61/SHINE and the latest flow performance studies were presented at this conference, see Ref.~\cite{cbm_poster}. 

\section{Acknowledgements}
\label{}
This work was partially supported by the Ministry of Science and Education of the Russian Federation, grant N 3.3380.2017/4.6, and by the National Research Nuclear University MEPhI in the framework of the Russian Academic Excellence Project (contract No. 02.a03.21.0005, 27.08.2013).
\bibliographystyle{elsarticle-num}
\bibliography{qm2018_na61_flow_proceedings.bib}
\end{document}